\documentclass[printer]{aa}  

\usepackage{graphicx}
\usepackage{natbib}
\usepackage{txfonts}
\begin{document}

   \title{Verification of the helioseismic Fourier-Legendre analysis for meridional flow measurements}


   \author{M. Roth
          \inst{1}
          \and
          H.-P. Doerr
          \inst{1}\thanks{Now at: Max-Planck-Institut f\"ur Sonnensystemforschung, Justus-von-Liebig-Weg 3, 37077 G\"ottingen, Germany}
          \and
          T. Hartlep\inst{2}\thanks{
Now at: Bay Area Environmental Research Institute, NASA Ames Research Center, Moffett Field, CA 94035, U.S.A.}
          }

   \institute{Kiepenheuer-Institut f\"ur Sonnenphysik, Sch\"oneckstr. 6, 79104 Freiburg, Germany\\
              \email{mroth@kis.uni-freiburg.de}
         \and
         Hansen Experimental Physics Laboratory, Stanford University, Standord, CA 94305, U.S.A.
         \email{thomas.hartlep@nasa.gov}
             }

   \date{Received September 15, 1996; accepted March 16, 1997}

 
  \abstract
  {Measuring the Sun's internal meridional flow is one of the key issues of helioseismology.
Using the Fourier-Legendre analysis is a technique for addressing this problem.} 
{We validate this technique with the help of artificial helioseismic data.}
{The analysed data set was obtained by numerically simulating the effect of the meridional flow on the seismic wave field in the full volume of the Sun. In this way, a 51.2-hour long time series was generated.
The resulting surface velocity field is then analyzed in various settings: Two $360^\circ \times 90^\circ$ halfspheres, two $120^\circ \times 60^\circ$ patches on the front and farside of the Sun (North and South, respectively) and two $120^\circ \times 60^\circ$ patches on the northern and southern frontside only. 
We compare two possible measurement setups: observations from Earth and from an additional spacecraft on the solar farside, and observations from Earth only, in which case the full information of the global solar oscillation wave field was available.
}
{We find that, with decreasing observing area, the accessible depth range decreases: the $360^\circ \times 90^\circ$ view allows us to probe the meridional flow almost to the bottom of the convection zone, while the $120^\circ \times 60^\circ$ view means only the outer layers can be probed.}
{These results confirm the validity of the Fourier-Legendre analysis technique for helioseismology of the meridional flow. Furthermore these flows are of special interest for missions like Solar Orbiter that promises to complement standard helioseismic measurements from the solar nearside with farside observations.}


   \maketitle
%

\section{Introduction}

Determining the Sun's internal meridional flow is one of the main challenges of solar physics. 
While the meridional flow amplitude at the solar surface is rather easy to measure from feature tracking~ (\cite{Woehl01,hathaway96}) and the established local helioseismic methods allow us to probe for the flow in shallow sub-surface layers (\cite{haber02,zhao04,irene10}), a significantly deeper helioseismic probing of the flow encounters difficulties. 
Since better knowledge about the structure of the meridional flow is desirable for input in to flux-transport dynamo models  (\cite{charbonneau10}), which are essential for understanding the Sun's magnetic activity cycle, it is the deep meridional flow that is of particular importance. Only recently, improved techniques of time-distance helioseismology~\citep{zhao13} and a new approach of global helioseismology~(\cite{schad13}) have provided first indications of the structure of the deep meridional flow.
However, today's helioseismic techniques rely on observations of the frontside of the Sun only. Within the coming years, this type of data might be complemented by missions like the Solar Orbiter, which provides seismic data from the solar farside.

In this paper we focus on inferring the Sun's internal meridional flow by using the Fourier-Legendre decomposition (FLD) analysis, which is a further development of the concept proposed by~\citet{braun98}. We employ numerical simulations to create artificial helioseismic data. These data are used to validate the FLD technique and to assess the accuracy of inferences that could be expected from different observational setups.

The paper is organized as follows: Section 2 describes the numerical simulations, including the flow model and the Fourier-Legendre technique as it is applied to various observational settings;
Section 3 presents the results; the conclusions are given in Section 4.

\section{Methods} 
      \label{Methods}
Our study rests on analysing artificial helioseismic data, i.e. a time series of a numerically obtained wavefield in the interior of the Sun, which is affected by a large-scale meridional flow.

\subsection{Numerical simulation} 
  \label{simulations}
The generation of artificial helioseismic data is identical to the method described in~\citet{hartlep2013}, which makes use of
a numerical code that solves the linearized propagation of helioseismic waves throughout the entire solar interior~(\cite{hartlepmansour2005}).
This code has been used in previous studies to simulate the effects of various localized sound-speed perturbations, e.g. for testing helioseismic farside imaging by simulating the effects of model sunspots on the acoustic field~(\cite{2008ApJ...689.1373H,2009SoPh..258..181I}), for validating time\,-distance helioseismic measurements of tachocline perturbations~(\cite{2009ApJ...702.1150Z}), and for studying the effect of localized subsurface perturbations (\cite{2011SoPh..268..321H}).

As described in~\citet{hartlep2013}, the code has been extended to include the effects of mass flows onto the propagation of helioseismic waves.
In short, the code models solar acoustic oscillations in a spherical domain by using the Euler equations that are linearized around a stationary background state.
As can be seen in~\citet{hartlep2013}, the equations are formulated in a non-rotating frame, and so rotation is accounted for by prescribing an appropriate flow.
This approach saves computing the usual Coriolis and centrifugal forces that appear in the equations for a rotating frame.

The excitation of acoustic waves, which is a non-linear process, is not included in the linearized equations. Instead, we mimic solar acoustic wave excitation by including in the momentum equation a random function that is only non-zero close to the solar surface.
Perturbations of the gravitational potential have been neglected and the adiabatic approximation has been used.
We only consider waves with constant entropy. This way, the entropy gradient of the background model does not enter
our equations and the model becomes convectively stable. 
Non-reflecting boundary conditions are applied at the upper boundary by means of an absorbing buffer layer with a damping coefficient that is zero in the interior and increases smoothly into the buffer layer. A small amount of viscous damping was added for stability reasons.

The simulation in this study resolves spherical harmonics of angular degree between 0 to 170.
A staggered Yee scheme (\cite{1966ITAP...14..302Y}) is used for time integration, with a time step of one second.

\subsection{Flow model}
For the background flow, we have imposed a stationary model of the solar meridional circulation according to \citet{hartlep2013}. The flow model described there is based on the flow model in \citet{rempel2006}.
The meridional flow is a single-cell circulation in each of the meridional quadrants.
As described in \citet{hartlep2013}, the amplitude 
was chosen so that the resulting meridional flow has a maximum velocity of 500 m/s. 

But, since these simulations are linear, this kind of an increase does not change the physics but merely increases the signal-to-noise ratio.
Hence, a simulated time series of 51.2 h length has the same signal-to-noise ratio as a 3.65 year-long time series with realistic speeds.


\subsection{Fourier-Legendre decomposition} 
  \label{FLD}
The FLD method is the extension to spherical geometry of the Fourier-Hankel decomposition (FHD) which has been used in the past to study wave absorption in sunspots ~\citep{braun88}. It is a helioseismic technique that is also suited for the measurement of the meridional flow~\citep{braun98}. Because the FLD method is sensitive to low-degree oscillation modes,
it promises probing the average meridional flow in deep layers.
An early version of this FLD analysis pipeline was tested on GONG (Global Oscillation Network Group) data by~\citet{doerr2010}, who successfully compared near-surface measurements of the meridional flow with results obtained by ring-diagram analyis.

The time dependent, two-dimensional oscillation signal on the solar surface $V(\theta,\phi,t)$ can be expressed in terms of a superposition of two wavefields travelling pole- and equatorwards, respectively:
\begin{equation}
V(\theta,\phi,t)=\sum_{nlm} \left[ A_{lm}H_m^{(1)}(L\theta)+B_{lm}H_m^{(2)}(L\theta)\right]{\rm e}^{{\rm i}m\phi+\omega_{nl}t}
\end{equation}
with values of $m\approx 0$. Here $\theta$ denotes the
co-latitude and $\phi$ the longitude on the solar surface, $\omega_{nl}$ is the angular frequency of a mode with harmonic degree $l$, $L=[l(l+1)]^{1/2}$, and azimuthal order $m$. The quantitities $H_m^{(1,2)}$ are travelling wave associated Legendre functions, described in analogy to \cite{Nussenzweig1965}:\begin{equation}
H_{m}^{(1,2)}= (-1)^m \frac{(l-m)!}{(l+m)!}\left[P_l^m(\cos\theta \pm \frac{2\rm i}{\pi}Q_l^m(\cos\theta)\right]\ ,
\end{equation}
where $P_l^m$ and $Q_l^m$ are associated Legendre functions of first and second kind, respectively.


Due to advection, a meridional flow will result in a frequency shift $\Delta\nu_{nl}$ 
between the poleward and equatorward components~\citep{braun98,goughtoomre83} 
\begin{equation}
\Delta\nu_{nl}=\frac{l}{\pi R_\odot}\int \bar{U}_{\rm mer}(r) K_{nl}(r)\, \mathrm{d}r\ ,
\end{equation}
where $\bar{U}_{\rm mer}(r)$ is the averaged meridional flow over the observed region of interest as a function
of the position $r$ inside the Sun. The sensitivity kernel $K_{nl}(r)$ is the kinetic energy density of a given
mode $(n,l)$ which is also a function of the position $r$ inside the Sun.

This frequency shift $\Delta\nu_{nl}$ can be measured from the power spectra of the two series, $A_{lm}$ and $B_{lm}$, by fitting asymmetric Lorentzian profiles to the individual peaks present. 
The guess frequencies for the fits are obtained from the standard solar {\em Model S}~(\cite{jcd}).  We then employ the SOLA (subtractive optimally localized averages)
inversion technique~(\cite{pijpers94}) to construct localized average inversion kernels at a given depth $r$.

\subsection{Data analysis}
The simulation results have been tracked with the mean rotation, i.e. transformed to a corotating reference frame with the mean rotation period. The surface velocity field was projected on a heliographic grid and the region of
interest is cut out. We study three different observational scenarios:
\begin{enumerate}
\item The northern and southern hemispheres, i.e. two patches with angular dimensions of $360^\circ\times 90^\circ$ in longitude and latitude, respectively. This corresponds to an ideal observational setup, where data from the whole solar surface is available for helioseismic analysis.


\item Four patches, each with the dimensions $120^\circ\times 60^\circ$. Two of the patches are on the northern and southern frontside, and two are $180^\circ$ in longitude apart on the farside, respectively. The patches are centred at $\pm 35^\circ$ latitude, respectively. This corresponds to combined observations in the ecliptic from, e.g. Earth and from a spacecraft on the farside of the Sun, as  might be possible with missions like the Solar Orbiter~\citep{mueller2013, roth2007}.

\item Two patches on the frontside only, with the dimensions $120^\circ\times 60^\circ$. The patches are located at $\pm 35^\circ$ latitude. This setup corresponds to the standard case of today's helioseismic observations with one viewpoint only.

\end{enumerate}

For all three cases, the artificial Dopplergrams are fed into the Fourier-Legendre decomposition procedure as described above.
This yields time series $A_{lm}(t)$ and $B_{lm}(t)$, from which power spectra $P_{lm}^{(A)}$ and $P_{lm}^{(B)}$ of north- and southwards travelling waves are generated for each hemisphere, respectively. These power spectra are then averaged over the range of azimuthal orders $m=-25,\dots,25$. This is meant to include a sufficient number of modes that are affected by the meridional flow, as they have a pole- or equatorward directed component of the wave vector. Then, the m-averaged power spectra are smoothed by applying a running boxcar window over three frequency bins.
The peaks in the averaged power spectra are then fitted using asymmetric Lorentzian profiles to determine the individual central mode frequencies $\nu_{nl}^{(A)}$ and $\nu_{nl}^{(B)}$. Finally, frequency differences $\delta\nu_{nl}=\nu_{nl}^{(A)}-\nu_{nl}^{(B)}$ are calculated.
The set of frequency differences is then inverted for the meridional flow, as described above.

\section{Results} 
      \label{Results}
      
      \subsection{Meridional flow measurements}
\begin{figure*}
   \centering
    \includegraphics[width=\columnwidth]{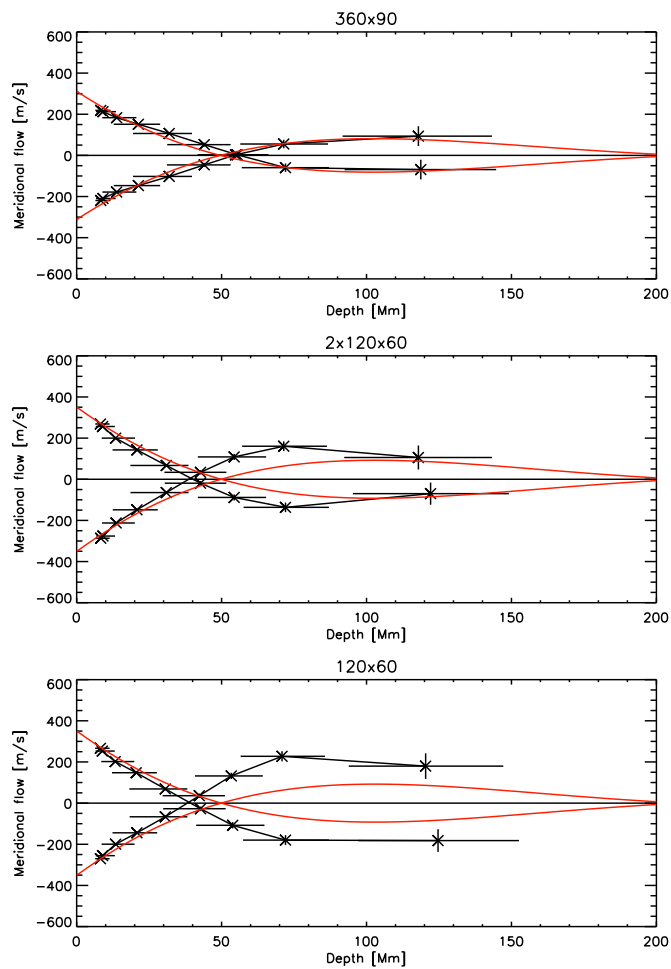}
\caption{Inversion results for the three observational setups. First row: Meridional flow measurements with both hemispheres of the Sun being observed. Second row: Same as top, but with two patches for each hemisphere observed on the front and on the farside of the  Sun. Third row: Same as top but with observations of two patches, one on the northern, the other on the southern hemisphere. The red lines indicate the averaged flow amplitudes below the observed patches on the northern and southern hemispheres, as taken from the input model. Horizontal error bars represent the quantile width of the averaging kernels.}
\label{fig1}
    \end{figure*}

Figure~\ref{fig1} displays the main results of this study. In the first case, where both hemispheres of the Sun could be observed, the input meridional flow can be well recovered (Fig.~\ref{fig1} (top)) down to a depth of 110 Mm. This case corresponds to an ideal observational setup, i.e. the oscillation signal is available for the full solar globe.

In the second case, where only part of the solar surface could be observed, i.e. two patches per hemisphere, one on the front- the other on the farside of the Sun, the flow could still be recovered well. The inversion result agrees well with the input model down to about 30 Mm below the solar surface. Below that, the inversion result deviates significantly from the input flow.

In the third case, where only frontside observations are available, the flow can be reliably recovered down to only 50 Mm. At deeper layers the inversion results deviate strongly from the input flow. 

We note that, since the numerical data included only low degree modes, the surface flow was not recovered in
all settings. The flow measurements were possible at depths greater than 10 Mm.

\subsection{Power leaks}
According to theoretical forward calculations by \cite{roth2008} and \cite{goughhindman}, the frequency shifts introduced by a meridional flow should be negligible.  Given the results by \cite{goughhindman}, the meridional flow should introduce a power redistribution among affected modes. In fact, this power redistribution comes from the coupling of modes owing to advection, which affects the eigenfunctions~(\cite{lavely,schad2011}). This is an effect that is evaluated in global helioseismology to measure the meridional flow~(\cite{schad13}. 

\begin{figure}
\centering
\includegraphics[width=\columnwidth]{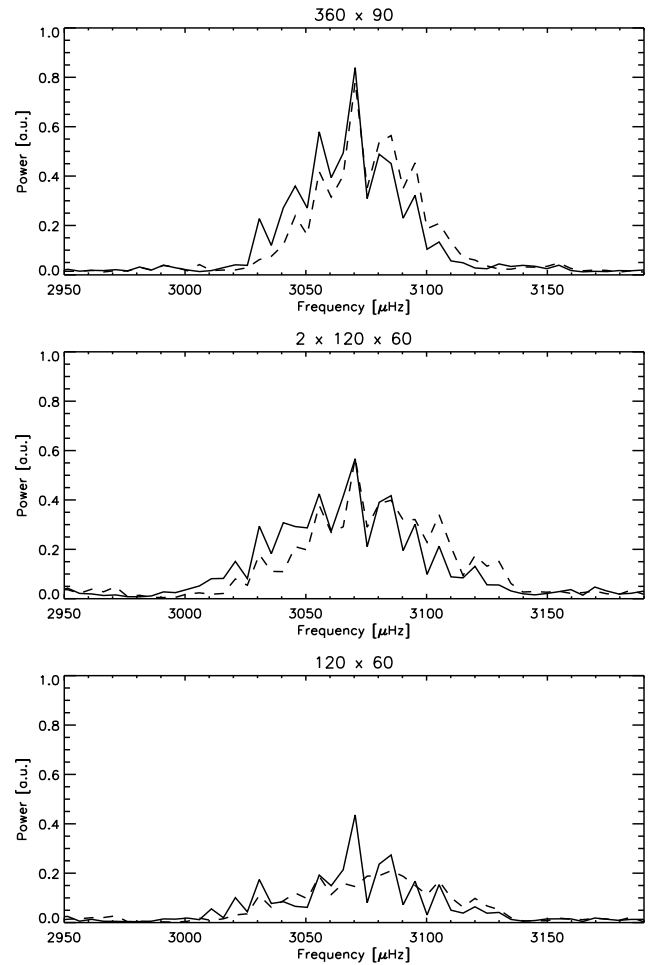}
\caption{Exemplary power spectra averaged over azimuthal orders $m=-25,\dots,25$ for the poleward (solid) and equatorward (dashed) travelling waves with harmonic degree $l=90$ and radial order $n=7$. From top to bottom, the viewing area decreases from $360^\circ \times 90^\circ$, over $2\times 120^\circ \times 60^\circ$, to $120^\circ \times 60^\circ$. Power is normalized with the same factor in all three plots.}
\label{fig2}
\end{figure}

Nevertheless, in this study, we successfully make us of measurable frequency shifts. 
By studying the power spectra of pole- and equatorwards travelling waves in detail, it becomes obvious that the modes are actually not measurably shifted and that this frequency shift must be spurious. Two effects need to be considered. First, due
to the fact that only parts of the solar surface are used for Fourier-Legendre decomposition, the spherical harmonics are no longer orthogonal on these reduced areas of the solar surface. As a result, spatial leaks appear.
 Fig.~\ref{fig2} shows the power spectra for our three observational setups.
As visible in Fig.~\ref{fig2}, this number of leaks increases with decreasing observing area, i.e. the central peak amplitude decreases and power leaks into a broad range of side lobes.
Second, the coupling of modes, which are caused by advection, introduces a number of side lobes i.e. power leaks, in the mode spectra, too. 
By comparing pole- and equatorward propagating waves, it can be seen that the individual central mode peak and its leaks do not change their position on the frequency axis. But, owing to the presence of the meridional flow, power is redistributed  among the leaks in a way that the weight of the whole set of coupled modes is shifted in frequency.
Consequently, smoothing the power spectra over frequency and fitting a common asymmetric Lorentzian profile to the resulting broad peak must yield shifted frequencies. This apparent frequency difference between pole- and equatorward travelling waves can then be employed for inversions. Making use of the mode kinetic energy, however, needs to be considered as an approximation to proper sensitivity kernels, which are not yet available  for this technique.

\section{Conclusion} 
      \label{conclusion}
We have validated the Fourier-Legendre decomposition technique as a means of measuring the meridional flow. For this we employed artificial helioseismic data.
The technique was applied to three settings that mimic observation scenarios. These range from ideal conditions of full surface observations of the oscillation signal to the grade of scenarios, which are more compatible with observations that are technically possible today, i.e. observing parts of the solar hemisphere on the front- and farside only.  

As a main result, we were able to demonstrate that the technique successfully recovers the meridional flow that was included in the simulation. This works  better when more of the solar surface is available for helioseismic analysis. We conclude that in the case that one hemisphere can be fully observed, a meridional flow within of 20 m/s amplitude could be inferred by a 3.65-year long time series with this technique, if the results from the 51.2 hours of artificial data are comparable to 3.65 years of observational data, which is not, a priori, clear, especially as helioseismic techniques work very well on simulations that are linear or convectively stable. More advanced simulations should therefore be considered in future for testing this technique, too.

We note that the technique makes use of an apparent frequency shift, which comes from the fact that power is redistributed between modes that are coupled by the meridional flow. This creates asymmetric weights in the power distribution of a mode and its leaks that are close-by in frequency, which can be interpreted as a frequency shift. Therefore, we can confirm the theoretical results obtained by \cite{goughhindman}, but are still  able to use the technique to measure the meridional flow down to great depths.  This works as long as the individual power leaks are not resolved in frequency because then the mode kinetic energy can serve as an approximation of the proper sensitivity kernels for this technique. The derivation of these kernels would be part of a future study. Thus, making use of the power redistribution would therefore   be the next step in developing this technique further because, accounting for the leakage of modes promises a better recovery of the flow at greater depths. As a consequence, the results of the FLD analysis will be more comparable to other methods, e.g. time-distance helioseismology. For example, ~\citet{jackiewicz2015} analyze this simulation too, using the time-distance technique and obtain more accurate inversion results in deeper layers.
Still, the method promises advantages in its capability of easy incorporation of data from multiple vantage points, which is a result from the applicability of the method to various patch geometries, in contrast to the ring-diagram technique.

Concerning multiple vantage points, we conclude that the meridional flow could be detected down to a large percentage of the convection zone using the Fourier-Legendre analysis technique when observational data from front- and farside observations are combined. Therefore our study may be interesting when planning upcoming space missions, e.g. Solar Orbiter.

\begin{acknowledgements}
M.R. thanks D. Braun for useful discussions. The authors thank the unknown referee for valuable comments on the manuscript.
The research leading to these results has received funding from the European
Research Council under the European Union's Seventh Framework Program (FP/2007-2013) /
ERC Grant Agreement no. 307117.
\end{acknowledgements}

\bibliographystyle{aa}

\bibliography{mroth_bib}

\end{document}